\documentclass[aps, superscriptaddress, twocolumn, prb]{revtex4-2}
\usepackage{amsfonts}
\usepackage{amssymb}
\usepackage{amsmath}
\usepackage{graphics}
\usepackage{graphicx}
\usepackage{subfigure}
\usepackage{dcolumn}

\usepackage{bm}
\usepackage{color}
\usepackage{epstopdf}

\definecolor{Gray}{gray}{0.1}

\DeclareMathAlphabet\mathbfcal{OMS}{cmsy}{b}{n}

\begin{document}

\title{Intense high-order harmonic generation in giant fullerene molecule C$%
_{240}$}
\author{H.K. Avetissian}
\affiliation{Centre of Strong Fields Physics at Research Institute of Physics, Yerevan State University,
Yerevan 0025, Armenia}
\author{S. Sukiasyan}
\affiliation{Centre of Strong Fields Physics at Research Institute of Physics, Yerevan State University,
Yerevan 0025, Armenia}
\affiliation{Max-Planck-Institut f\"ur Kernphysik, Saupfercheckweg 1, 69117 Heidelberg, Germany}
\author{T.M. Markosyan}
\affiliation{Centre of Strong Fields Physics at Research Institute of Physics, Yerevan State University,
Yerevan 0025, Armenia}
\author{G.F. Mkrtchian}
\thanks{Email: mkrtchian@ysu.am}
\affiliation{Centre of Strong Fields Physics at Research Institute of Physics, Yerevan State University,
Yerevan 0025, Armenia}

\begin{abstract}
In this work the extreme nonlinear optical response of a giant fullerene
molecule C$_{240}$ in strong laser field is studied. The investigation of
high-order harmonic generation in such quantum nanostructure is presented
modeling the C$_{240}$ molecule and its interaction with the laser field in
the scope of the tight-binding mean-field approach. Electron-electron
interaction is modeled by the parametrized Ohno potentail, which takes into
account long-range Coulomb interaction. The essential role of many body
Coulomb interaction in determining of harmonics intensities is demonstrated.
We also consider vacancy-deffected molecule C$_{240}$. The presence of a
single vacancy breaks the icosahedral symmetry leading to the emergence of
intense even-order harmonics. We examine the dependence of moderate
harmonics on laser frequency that shows the multiphoton resonant nature of
high harmonics generation. The dependence of cutoff harmonics on both laser
intensity and frequency are examined too.
\end{abstract}

\maketitle

\section{Introduction}

Intense light interaction with nanostructures can excite the electrons of
the system through multiphoton channels, leading to extreme nonequilibrium
states \cite{Avetissian2015relativistic}. The excited electrons subsequently
emit coherent electromagnetic radiation, encompassing tens to hundreds of
harmonics of the incident light \cite%
{agostini2004physics,kohler2012frontiers}. This fundamental process in
intense laser-matter interaction is known as high harmonic generation (HHG)
phenomenon \cite{corkum1993plasma,lewenstein1994theory}. In atoms, HHG has
been widely used to produce coherent extreme ultraviolet radiation, allowing
access to the extreme time resolution of the underlying quantum processes
and enabling attosecond physics \cite%
{corkum2007attosecond,krausz2009attosecond}. Among the diverse range of
nanostructured materials suitable for nonlinear extrime optical
applications, carbon allotropes hold a central position \cite%
{falcao2007carbon,tiwari2016magical}. One of the carbon allotropes are
fullerenes \cite{smalley1997discovering} which are large molecules formed by
closing a graphite sheet, where the required curvature is achieved by
incorporating twelve pentagons among a given number of graphene hexagons.
The most well-known fullerene is the buckminsterfullerene C$_{60}$ \cite%
{kroto1985c60}, which possesses icosahedral symmetry. The discovery of
fullerene C$_{60}$ through laser evaporation of graphite was triggered the
study of many other fullerene molecules. Larger fullerenes, often referred
to as giant fullerenes, can also be constructed with icosahedral symmetry 
\cite{kroto1988formation}. These large fullerenes can be visualized as
cut-out pieces of graphene that are folded into an icosahedron.
Consequently, they exhibit similar properties to graphene \cite%
{geim2009graphene} or graphene quantum dots \cite{gucclu2014graphene}, while
remaining stable due to their closed topological structure. Note that in
continuous limit C$_{60}$ and related molecules are well described by the
Dirac equation in the curved space and in the field of a monopole \cite%
{gonzalez1993electronic,gonzalez1992continuum}. Giant or large fullerenes
have been the subject of active research since the 1990s. For a more
comprehensive overview, we refer the reader to references\cite%
{york1994density,scuseria1995equilibrium,scuseria1996ab,itoh1996structure,xu1996n,haddon1997c240,heggie1998quantitative}
for earlier studies and references \cite%
{dunlap2006efficient,zope2008static,calaminici2009first,dunk2012closed,martin2017giant,wang2020structural,ghavanloo2023experimental}
for more recent investigations.

In the field of HHG, enhancing conversion efficiency is of utmost
importance. This efficiency strongly relies on the density of emitters and
the density of states of these emitters. To this end, molecular systems,
clusters, and crystals have shown potential in significantly increasing
harmonic intensity compared to atomic systems, as they can exploit multiple
excitation channels \cite{donnelly1996high,vozzi2005cluster,smirnova2009high}%
. As a result, there has been a growing interest in extending HHG to
carbon-based materials, such as semimetallic graphene \cite%
{mikhailov2008nonlinear,avetissian2012creation,avetissian2013multiphoton,bowlan2014ultrafast,al2014high,chizhova2016nonlinear,avetissian2016coherent,dimitrovski2017high,avetissian2018impact,sato2021high,zurron2019optical,mrudul2021high,zhang2021orientation,dong2021ellipticity,avetissian2022efficient,murakami2022doping,tamaya2023shear}%
, graphene quantum dots \cite%
{JETP2022high,JN2022high,JETPL2022laser,gnawali2022ultrafast}, and
fullerenes \cite%
{bauer2001c,zhang2005optical,zhang2006ellipticity,ciappina2008high,ganeev2009higher,ganeev2009high,redkin2010simulation,topcu2019drastically,ganeev2013high,zhang2020high,avetissian2021high,avetissian2023disorder}%
. . Experimental studies, namely Refs. \cite{ganeev2009higher,ganeev2009high}%
, have reported a robust harmonic signal from C$_{60}$ plasma. Additionally,
theoretical works have predicted strong HHG from both C$_{60}$ \cite%
{zhang2005optical,zhang2006ellipticity,avetissian2021high,avetissian2023disorder}
and C$_{70}$ molecules \cite{avetissian2021high} and solid C$_{60}$ \cite%
{zhang2020high}. Notably, the increase in conducting electrons in fullerene
molecules leads to a subsequent rise in density of states, thereby opening
up new channels that can amplify the HHG signal. Consequently, exploring the
HHG process in giant fullerenes becomes a compelling area of interest. With
the increasing fullerene size, the molecules are subject to various types of
defects. Therefore, investigating the impact of defects on HHG in large
fullerenes holds significance. Recent research involve effects of disorder,
impurities, and vacancies on HHG in solids \cite%
{orlando2018high,yu2019enhanced,yu2019high,pattanayak2020influence,iravani2020effects,chinzei2020disorder,hansen2022doping,xia2022theoretical,orlando2022ellipticity}%
. These studies have revealed that an imperfect lattice can enhance HHG
compared to a perfect lattice, especially when considering doping-type
impurities or disorders. For C$_{60}$ and C$_{180}$, it has been shown that
both diagonal and off-diagonal disorders break inversion symmetry, lift the
degeneracy of states, and create new channels for interband transitions,
resulting in enhanced high harmonic emission \cite{avetissian2023disorder}.
This raises intriguing questions about how vacancies specifically affect the
HHG spectra in large fullerenes. Vacancies can occur naturally or be
introduced in fullerenes through laser or ion/electron irradiation \cite%
{deng1993odd,banhart1999irradiation}. Taking into account that vacancy
defects introduce localized electronic states \cite{terrones2000coalescence}
and the HHG process is highly sensitive to electron wave functions, we can
expect new effects in the HHG process at consideration of vacancy-defected
fullerenes.

In this study, we present a microscopic theory that explores the extreme
nonlinear interaction of normal and single vacancy-defected fullerene C$%
_{240}$ with strong electromagnetic radiation. Particularly, we consider
coherent interaction with a linearly polarized electromagnetic radiation
taking into account collective electron-electron interactions. Employing the
dynamical Hartree-Fock approximation, we reveal the general and basal
structure of the HHG spectrum and its relation to molecular excitations and
icosahedral symmetry breaking of giant molecules.

The paper is organized as follows. In Sec. II, the model and the basic
equations are formulated. In Sec. III, we present the main results. Finally,
conclusions are given in Sec. IV.

\section{The model and theoretical approach}

We start by describing the model and theoretical approach. Fullerene
molecule C$_{240}$ and C$_{240}$ with a monovacancy is assumed to interact
with a mid-infrared or visible laser pulse that excites electron coherent
dynamics. For the brevity we refer vacancy-defected C$_{240}$ molecule as C$%
_{239}$. The schematic structure of these fullerene molecules are deployed
in Fig. 1. We assume a neutral molecules, which will be described in the
scope of the tight-binding (TB) theory. The electron-electron interaction
(EEI) is described in the extended Hubbard approximation \cite%
{martin1993coulomb,harigaya1994optical,avetissian2021high}. Hence, the total
Hamiltonian reads: 
\begin{equation}
\widehat{H}=\widehat{H}_{0}+\widehat{H}_{\mathrm{int}},  \label{H1}
\end{equation}%
where%
\begin{equation}
\widehat{H}_{0}=-\sum_{\left\langle i,j\right\rangle \sigma
}t_{ij}c_{i\sigma }^{\dagger }c_{j\sigma }+\frac{U}{2}\sum_{i\sigma
}n_{i\sigma }n_{i\overline{\sigma }}+\frac{1}{2}\sum_{i,j}V_{ij}n_{i}n_{j}
\label{Hfree}
\end{equation}%
is the free fullerene Hamiltonian. Here $c_{i\sigma }^{\dagger }$ creates an
electron with spin polarization $\sigma =\left\{ \uparrow ,\downarrow
\right\} $ at site $i$ ($\overline{\sigma }$ is the opposite to $\sigma $
spin polarization), and $\left\langle i,j\right\rangle $ runs over all the
first nearest-neighbor hopping sites with the hopping integral $t_{ij}$
between the nearest-neighbor atoms at$\ $positions $\mathbf{r}_{i}$ and $%
\mathbf{r}_{j}$. The density operator is: $n_{i\sigma }=c_{i\sigma
}^{\dagger }c_{i\sigma }$, and the total electron density for the site $i$\
is: $n_{i}=n_{i\uparrow }+n_{i\downarrow }$. The second and third terms in
Eq. (\ref{Hfree}) describe the EEI Hamiltonian, with the parameters $U$ and $%
V_{ij}$ representing the on-site, and the long-range Coulomb interactions,
respectively. The involved molecules contain single and double carbon bonds,
for which model Hamiltonian (\ref{Hfree}) has been parameterized extensively
over the years. The input Cartesian coordinates for C$_{240}$ are obtained
from the Yoshida database \cite{yoshida20vrml}. In the present paper, as
first approximation, monovacancy is simulated by removing one carbon atom.
The initial structures are further optimized with the help of IQmol programm 
\cite{gilbert2012iqmol}. Hence, in the vicinity of the vacancy the bond
lenghts are changed. There is also scenary when the structure undergoes a
bond reconstruction in the vicinity of the vacancy \cite{ding2005theoretical}%
. In either case, a local distortion of the lattice takes place resulting
states that are strongly localized around defects \cite%
{pereira2008modeling,lee2005diffusion}. For the one-electron hopping matrix
elements, which in this work have been restricted to the nearest neighbors,
we use values close to the graphene hopping matrix elements. The common
choice of hopping matrix element is $t_{0}=2.7$ eV, corresponding to the C-C
bond length of $d_{0}=1.42\mathrm{\mathring{A}}$, while for shorter or
longer bonds, its value is extrapolated using the linear relationship $%
t_{ij}=t_{0}+\alpha \left( d_{0}-\left\vert \mathbf{r}_{i}-\mathbf{r}%
_{j}\right\vert \right) $, with $\alpha =3.5\ \mathrm{eV/\mathring{A}}$
being the electron-phonon coupling constant. The EEI is modeled by the Ohno
potential \cite{ohno1964some}:

\begin{equation}
V_{ij}=\frac{U}{\sqrt{1+\frac{U^{2}\left\vert \mathbf{r}_{i}-\mathbf{r}%
_{j}\right\vert ^{2}}{V^{2}d_{m}^{2}}}},  \label{ohno}
\end{equation}%
where $V$ means the strength of the long range Coulomb interaction, and $%
d_{m}$ is the average bond length. Depending on the screening effects a
popular choice of parameters for the Coulomb interactions is $0\leq U\leq
4t_{0}$, and $V=0.5U$ \cite{harigaya1994optical,harigaya1998effects}.

The light-matter interaction is described in the length-gauge%
\begin{equation}
\widehat{H}_{\mathrm{int}}=e\sum_{i\sigma }\mathbf{r}_{i}\cdot \mathbf{E}%
\left( t\right) c_{i\sigma }^{\dagger }c_{i\sigma },  \label{Hint}
\end{equation}%
where $\mathbf{E}\left( t\right) =f\left( t\right) E_{0}\hat{\mathbf{e}}\cos
\omega t$ is the electric field strength, with the amplitude $E_{0}$,
frequency $\omega $, polarization $\hat{\mathbf{e}}$ unit vector, and pulse
envelope $f\left( t\right) =\sin ^{2}\left( \pi t/\mathcal{T}\right) $. The
pulse duration $\mathcal{T}$ is taken to be $10$ wave cycles: $\mathcal{T}%
=20\pi /\omega $. From the Heisenberg equation under the Hartree-Fock
approximation one can obtain evolutionary equations for the single-particle
density matrix $\rho _{ij}^{\left( \sigma \right) }=\left\langle c_{j\sigma
}^{\dagger }c_{i\sigma }\right\rangle $ \cite{avetissian2021high}: 
\begin{equation*}
i\hbar \frac{\partial \rho _{ij}^{\left( \sigma \right) }}{\partial t}%
=\sum_{k}\left( \tau _{kj\sigma }\rho _{ik}^{\left( \sigma \right) }-\tau
_{ik\sigma }\rho _{kj}^{\left( \sigma \right) }\right) +\left( V_{i\sigma
}-V_{j\sigma }\right) \rho _{ij}^{\left( \sigma \right) }
\end{equation*}

\begin{equation}
+e\mathbf{E}\left( t\right) \left( \mathbf{r}_{i}-\mathbf{r}_{j}\right) \rho
_{ij}^{\left( \sigma \right) }-i\hbar \gamma \left( \rho _{ij}^{\left(
\sigma \right) }-\rho _{0ij}^{\left( \sigma \right) }\right) ,
\label{evEqs1}
\end{equation}%
where $V_{i\sigma }$ and $\tau _{ij\sigma }$ are defined via$\ $density
matrix $\rho _{ij}^{\left( \sigma \right) }$ and its initial value: 
\begin{equation}
V_{i\sigma }=\sum_{j\alpha }V_{ij}\left( \rho _{jj}^{\left( \alpha \right)
}-\rho _{0jj}^{\left( \alpha \right) }\right) +U\left( \rho _{ii}^{\left( 
\overline{\sigma }\right) }-\rho _{0ii}^{\left( \overline{\sigma }\right)
}\right) ,  \label{V}
\end{equation}

\begin{equation}
\tau _{ij\sigma }=t_{ij}+V_{ij}\left( \rho _{ji}^{\left( \sigma \right)
}-\rho _{0ji}^{\left( \sigma \right) }\right) .  \label{tau}
\end{equation}%
In addition, we assumed that the system relaxes at a rate $\gamma $ to the
equilibrium $\rho _{0ij}^{\left( \sigma \right) }$ distribution. As we see,
due to the mean field modification hopping integrals (\ref{tau}) become
non-zero between the remote nodes, irrespective of the distance.

\begin{figure*}[tbp]
\includegraphics[width=0.75\textwidth]{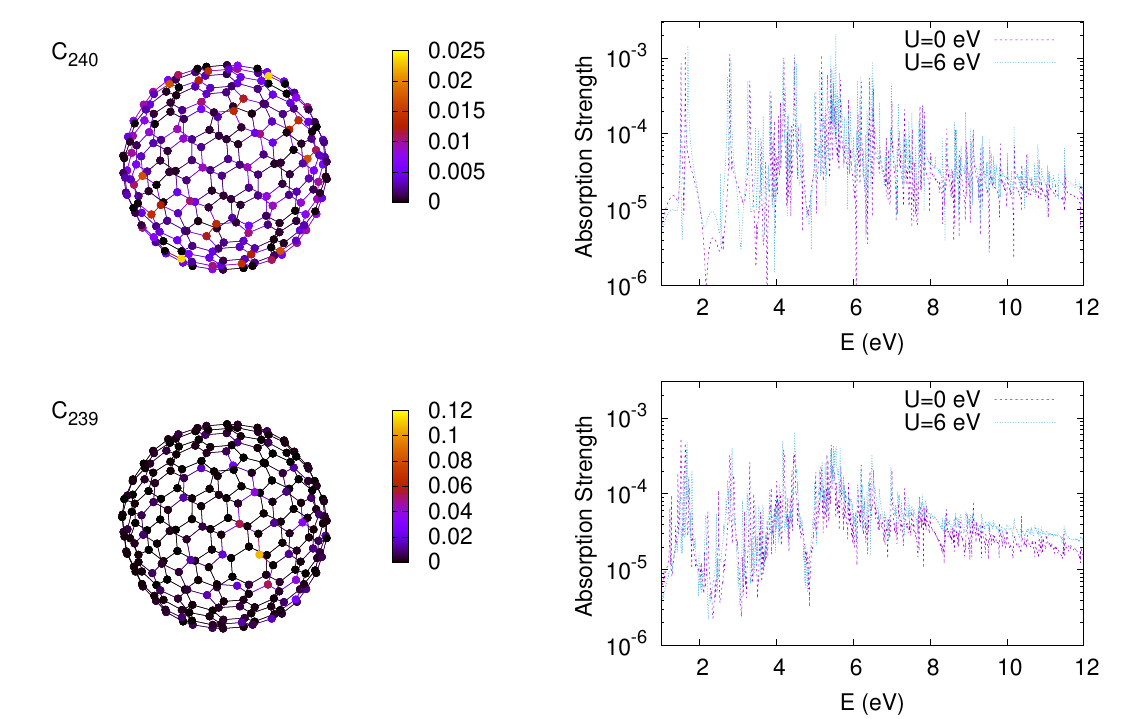}
\caption{The top and bottom pannels represent C$_{240}$ fullerene and C$%
_{240}$ with a monovacancy, respectively. For the brevity for refer to the
latter as C$_{239}$. Within each row, the following visualizations are
presented from left to right: electron probability density corresponding to
the highest energy level in the valence band on the 3D color mapped
molecular structures and the linear absorption spectra, for Coulomb
interaction turned on and off.}
\end{figure*}

\section{Results}

Now we discuss full numerical solution of the evolutionary equations for the
single-particle density matrix (\ref{evEqs1}) and to get more physical
insight, we study the question: which effects can be already observed in a
linear regime of interaction. The time propagation of Eq. (\ref{evEqs1}) is
performed by the 8-order Runge-Kutta algorithm. As an initial density matrix
we take a fully occupied valence band and a completely empty conduction
band. To study the HHG process in giant fullerene molecule we evaluate the
high-harmonic spectrum by Fourier transformation of the dipole acceleration, 
$\mathbf{a}\left( t\right) =d^{2}\mathbf{d(t)}/dt^{2}$, where the dipole
momentum is defined as $\mathbf{d}\left( t\right) =e\sum_{i\sigma }\mathbf{r}%
_{i}\rho _{ii}^{\left( \sigma \right) }\left( t\right) $: 
\begin{equation*}
\mathbf{a}\left( \Omega \right) =\int_{0}^{\mathcal{T}}\mathbf{a}\left(
t\right) e^{i\Omega t}W\left( t\right) dt,
\end{equation*}%
and $W\left( t\right) $ is the window function to suppress small
fluctuations \cite{zhang2018generating} and to decrease the overall
background (noise level) of the harmonic signal. As a window function we
take the pulse envelope $f\left( t\right) $. To obtain the mean picture
which does not depend on the orientation of the molecule with respect to
laser polarization, we take the wave polarization unity vector as $\hat{%
\mathbf{e}}=\left( 1/\sqrt{3},1/\sqrt{3},1/\sqrt{3}\right) $.

We begin by examining the effect of vacancy on the states near the Fermi
level. In Fig. 1, electron probability density corresponding to the highest
energy level in the valence band on the 3D color mapped molecular structures
are shown. As is seen from this figure, for a vacancy deffected case we have
state strongly localized around the vacancy. Thus, the presence of single
vacancy also breaks the icosahedral symmetry. To examine intrinsic molecular
transitions, we consider the extreme case of an external electric field that
has the shape of a delta-like impulse in time to excite all electronic
eigenmodes of the systems considered. In this case the relaxation rate is
taken to be very small $\hbar \gamma =0.5\ \mathrm{meV}$ to resolve
transitions as much as possible. The right pannels of Fig. 1 show linear
absorption spectra (in arbitrary units), for Coulomb interaction, turned on
and off. The peaks are intrinsic molecular excitation lines and the area of
a particular peak defines the weight of the oscillator strengths. The
effects of the EEI are similar to those of the fullerene molecule $C_{60}$
molecule \cite{harigaya1994optical}. The Coulomb interaction shift peaks to
higher energies, and oscillator strengths at higher energies have relatively
larger weight than in the free electron case. These effects are due to the
fact that the long range Coulomb interactions (\ref{ohno}) give rise to
large hopping integrals between the remote nodes (\ref{tau}) in the
Hartree-Fock approximation. For the vacancy defected case the transitions
are overall suppresed compared to intriscic case, although the low energy
transitions are strongly modified. From this figure we also see that the
optical gap in fullerene molecule $\mathrm{C}_{240}$ is approximately $1.7\ 
\mathrm{eV}$, which is narrower than that in $\mathrm{C}_{60}$ ($2.8\ 
\mathrm{eV}$). Notably, in both cases the absorption spectra exhibit many
peaks up to the high energies, suggesting the presence of efficient
multiphoton excitation channels and subsequent high-energy single-photon
transitions. These factors play a significant role in shaping the HHG
spectrum, as we will explore in the following.

\begin{figure}[tbp]
\includegraphics[width=0.42\textwidth]{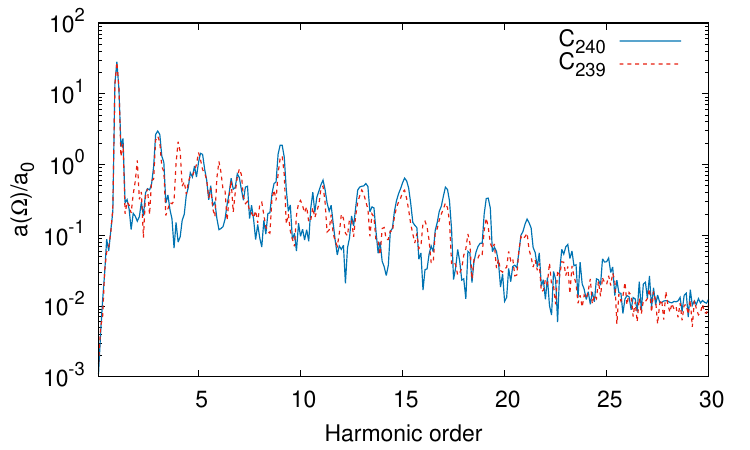}
\caption{The HHG spectra in the strong-field regime in logarithmic scale via
the normalized dipole acceleration Fourier transformation $a\left( \Omega
\right) /a_{0}$ (in arbitrary units) for C$_{240}$ and for C$_{239}$. The
laser frequency is $\protect\omega =1.2\ \mathrm{eV}/\hbar $. The spectra
are shown for EEI energy $U=6$ $\ \mathrm{eV}$.}
\end{figure}

Next, we will study more comprehensive the extreme nonlinear response of
giant fullerene molecule C$_{240}$ and its vacancy-defected conterpart C$%
_{239}$. For all further calculations, except of \ Fig. 7, the relaxation
rate is taken to be $\hbar \gamma =0.1\ \mathrm{eV}$. For the convenience,
we normalize the dipole acceleration by the factor $a_{0}=e\overline{\omega }%
^{2}\overline{d},$ where $\overline{\omega }=1\ \mathrm{eV}/\hbar $ and $%
\overline{d}=1\ \mathrm{\mathring{A}}$. The power radiated at the given
frequency is proportional to $\left\vert \mathbf{a}\left( \Omega \right)
\right\vert ^{2}$.

In Fig. 2, we show the typical HHG spectra in the strong field regime ($%
E_{0}=0.5\ \mathrm{V/A}$) for both molecules. For the C$_{240}$ molecule,
the presence of inversion symmetry restricts the appearance of only odd
harmonics in the HHG spectrum. In contrast, the introduction of a single
vacancy in the C$_{239}$ molecule disrupts its icosahedral symmetry,
resulting in the prominent emergence of even-order harmonics with enhanced
intensity. Besides, we see strongly nonlinear picture, where the strength of
the $9$th harmonic surpasses that of the $5$th and $7$th harmonics.
Additionally, a distinctive plateau spanning from the $11$th to the $21$st
harmonics exhibits comparable strengths. Notably, for the C$_{239}$
molecule, the harmonics near the cutoff display a slight suppression
relative to C$_{240}$ one. This disparity is attributed to the differing
effectiveness of excitation channels, which favors enhanced harmonics in the
case of C$_{240}$ molecule (see Fig. 1). 
\begin{figure}[tbp]
\includegraphics[width=0.42\textwidth]{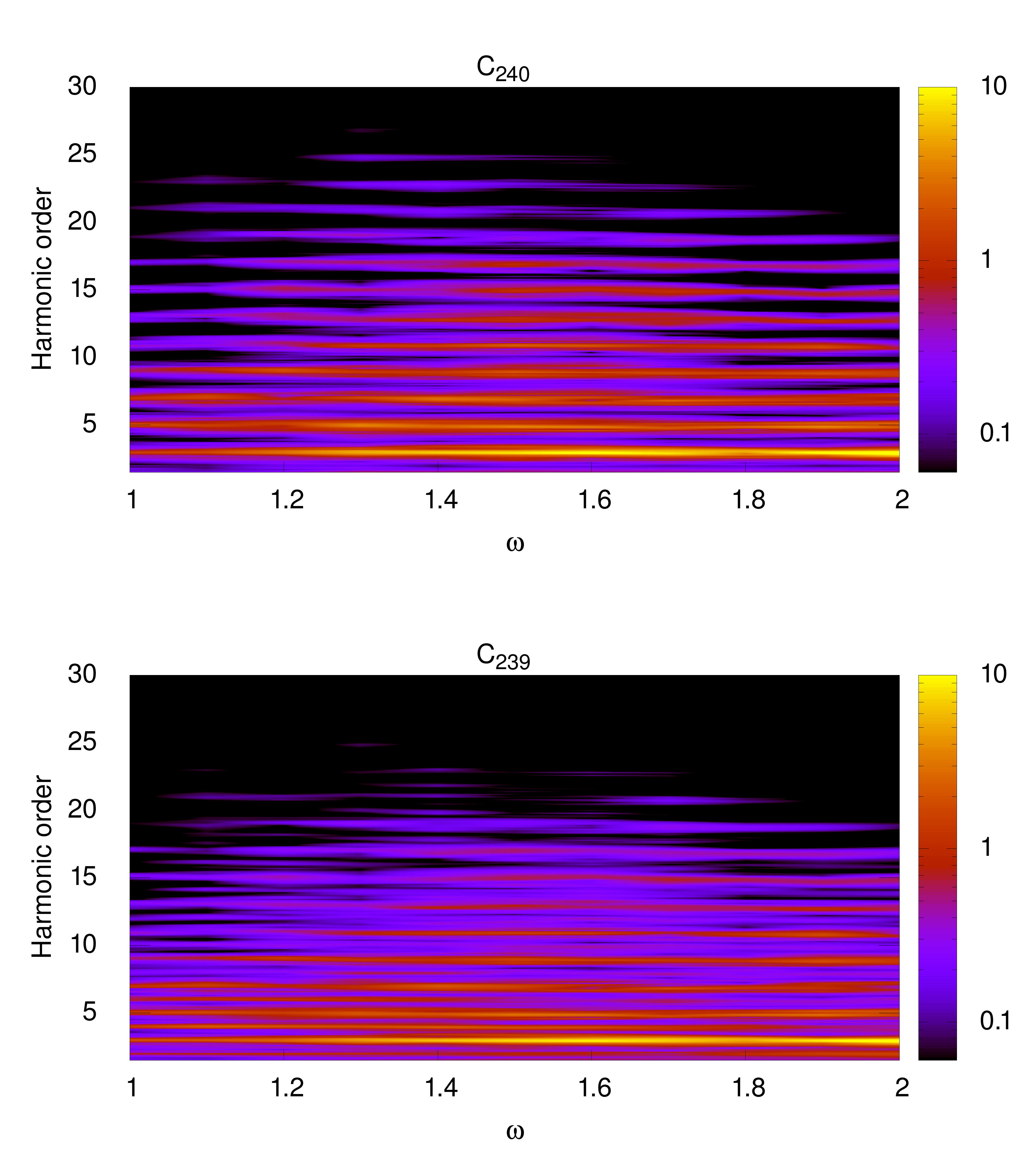}
\caption{The dependence of the HHG spectra on the wave field frequency is
illustrated for C$_{240}$ (top) and C$_{239}$ (bottom) using the normalized
dipole acceleration Fourier transformation, $a\left( \Omega \right) /a_{0}$,
plotted on a logarithmic scale. The wave amplitude is taken to be $%
E_{0}=0.5\ \mathrm{V/\mathring{A}}$. The relaxation rate is set to $\hbar 
\protect\gamma =0.1\ \mathrm{eV}$. The EEI energy is U = 6 eV.}
\end{figure}

Let us now consider the influence of the pump wave frequency on the HHG
process within the energy range of $\hbar \omega =1-2$ $\ \mathrm{eV}$. This
analysis is presented in Fig. 3 that illustrates the frequency-dependent HHG
spectra. Notably, we discern that the position of the cutoff harmonic $N_{%
\mathrm{cut}}$ demonstrates a relatively gradual response to changes in the
wave field of frequency $\omega $. Additionally, this cutoff exhibits
distinctive peaks within the mid-frequency range. It's worth noting that in
atomic HHG processes involving free continua, the cutoff harmonic position $%
N_{\mathrm{cut}}\sim \omega ^{-3}$ \cite{lewenstein1994theory}. Furthermore,
a noteworthy feature emerges when considering the C$_{239}$ molecule:
even-order harmonics are suppressed for higher frequency pump waves. This
phenomenon can be attributed to the fact that with higher frequency pump
waves, excitation and recombination channels predominantly involve highly
excited states that still retain the inversion symmetry. Of particular
interest is the plateau region within the spectra. Here, a pattern of
alternating variation in relation to frequency becomes evident, a hallmark
of multiphoton resonant transitions between the valence and conduction
bands. This resonant behavior is further illuminated by Figs. 4 and 5, where
we visualize the dependency of emission strength for the preplateau
harmonics on the pump wave frequency. It is apparent that these harmonics
exhibit resonant behavior. Upon a closer examination of Fig. 1, we discern
that the molecular excitations exhibit peaks coinciding with these resonant
frequencies, providing supplementary evidence for the multiphoton resonant
transitions. For instance, in the case of molecule C$_{240}$, the highest
peak for the $5$th harmonic emerges at around $1.3\mathrm{eV}$. This
frequency aligns with the local peak at $5\omega \sim 6.5\mathrm{eV}/\hbar $
in Fig. 1, accompanied by multiple excitation channels. Similarly,
considering molecule C$_{239}$, the peak for the $6$th harmonic is proximate
to $1.18$ $\ \mathrm{eV}$, in acordance with the local peak at $6\omega \sim
7\mathrm{eV}/\hbar $ in Fig. 1. The peaks displayed in Figs. 4 and 5
correspond with similar peaks in the molecular excitation spectra, as
depicted in Fig. 1. 
\begin{figure}[tbp]
\includegraphics[width=0.4\textwidth]{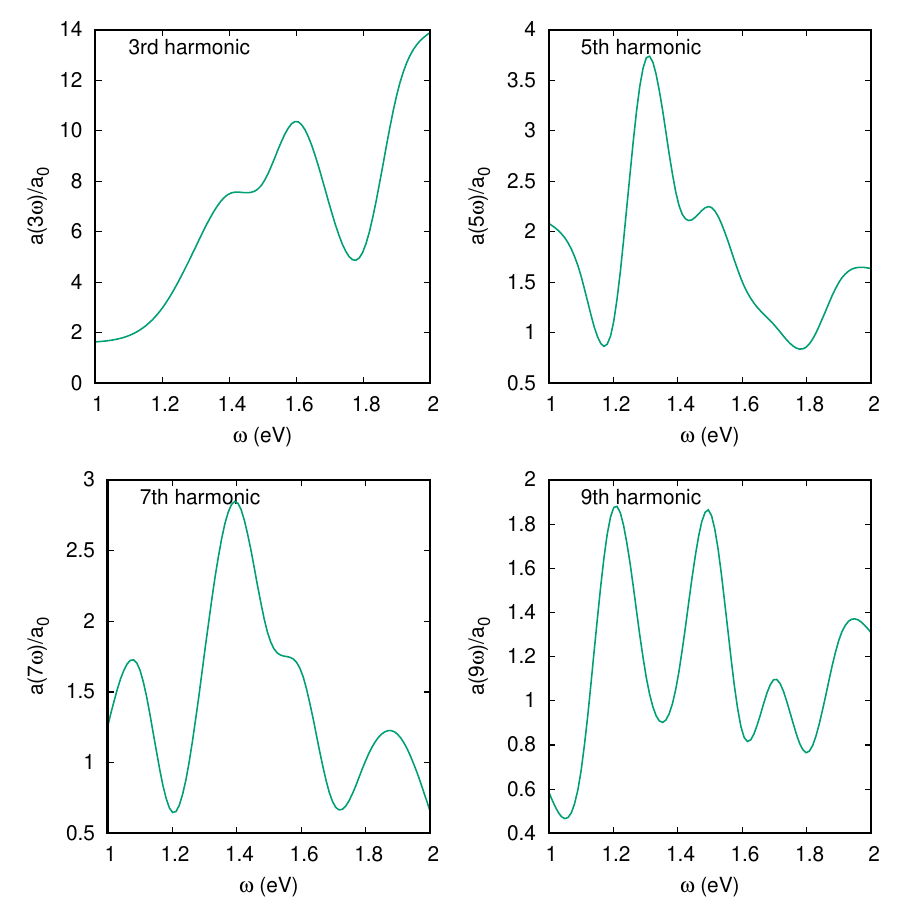}
\caption{The dependence of emission strength in the case of C$_{240}$ for
the 3rd, 5th, 7th, and 9th harmonics on the pump wave frequency for the
setup of Fig. 3.}
\end{figure}
\begin{figure}[tbp]
\includegraphics[width=0.4\textwidth]{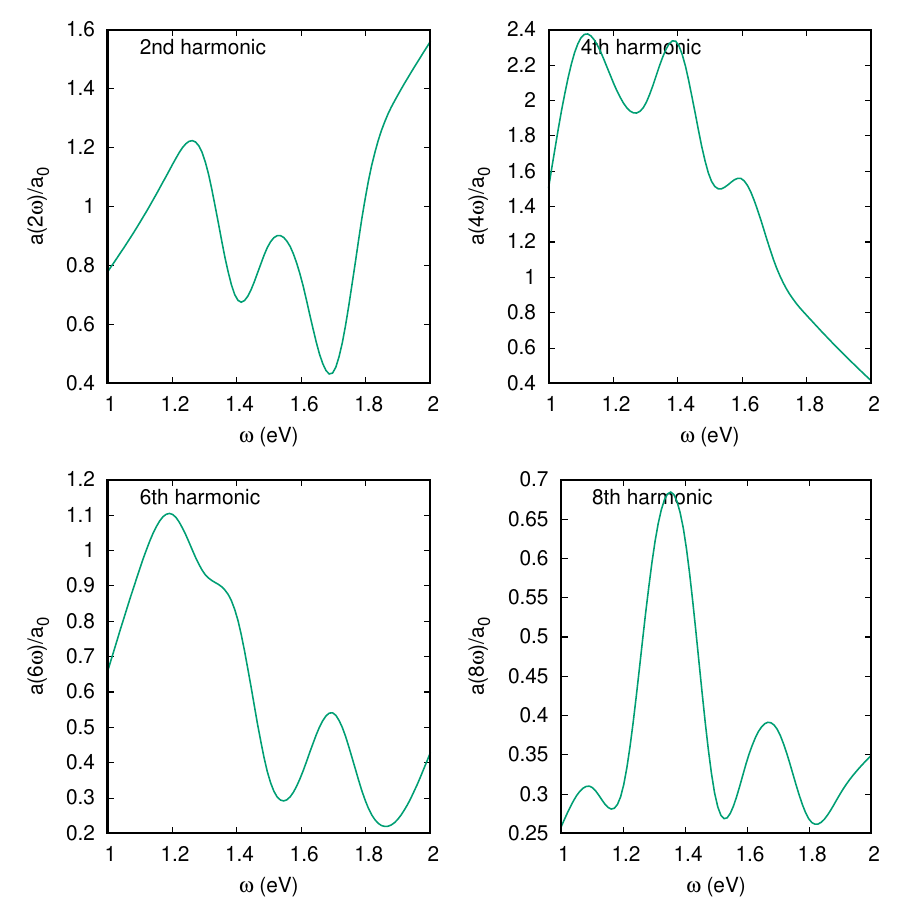}
\caption{The dependence of emission strength in the case of C$_{239}$ for
the 2nd, 4th, 6th, and 8th harmonics on the pump wave frequency for the
setup of Fig. 3.}
\end{figure}

The multiphoton resonance-driven characteristics are further supported by
the evident alteration in the population of energy levels within the valence
and conduction bands, as highlighted in Fig. 6. This figure presents the
post-interaction population distribution of energy levels, demonstrating a
marked departure from the equilibrium distribution. This discrepancy
underscores the substantial impact of multiphoton resonant transitions
within the HHG process of giant fullerene C$_{240}$ under the influence of
intense near-infrared laser fields.

Continuing our exploration, let us examine the influence of the relaxation
rate on the HHG phenomenon across a span of $\hbar \gamma =0.1-0.2$ $\ 
\mathrm{eV}$. The corresponding dependencies of the HHG spectra on the
relaxation rate are presented in Fig. 7. It is discernible that HHG exhibits
resistance to relaxation processes, with preplateau harmonics, in
particular, displaying notable robustness.

\begin{figure}[tbp]
\includegraphics[width=0.4\textwidth]{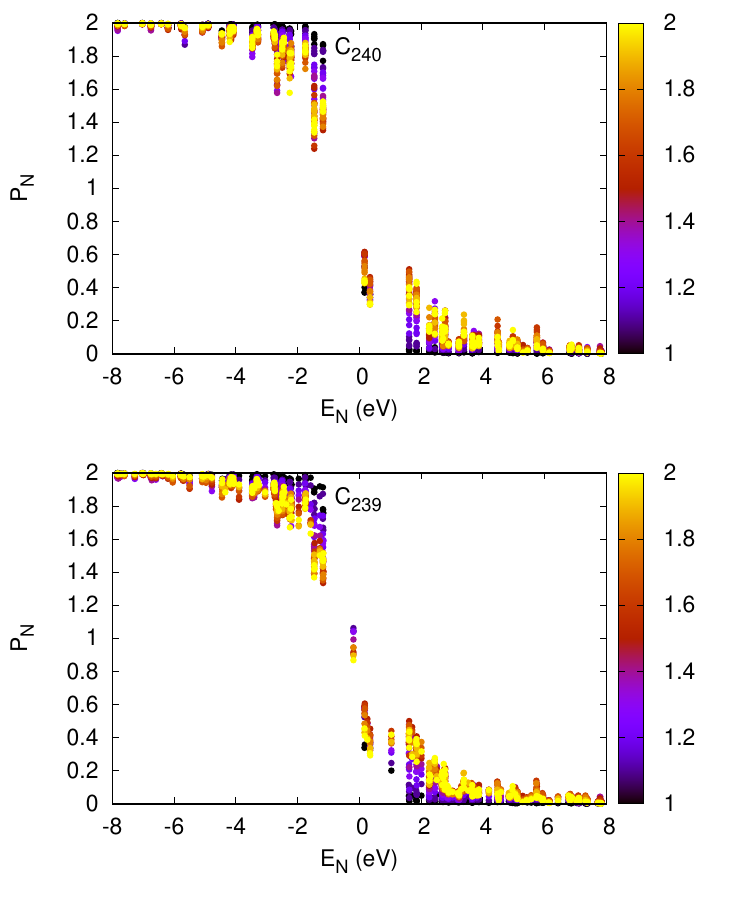}
\caption{The residual population of levels for the setup of Fig. 3.}
\end{figure}

\begin{figure}[tbp]
\includegraphics[width=0.4\textwidth]{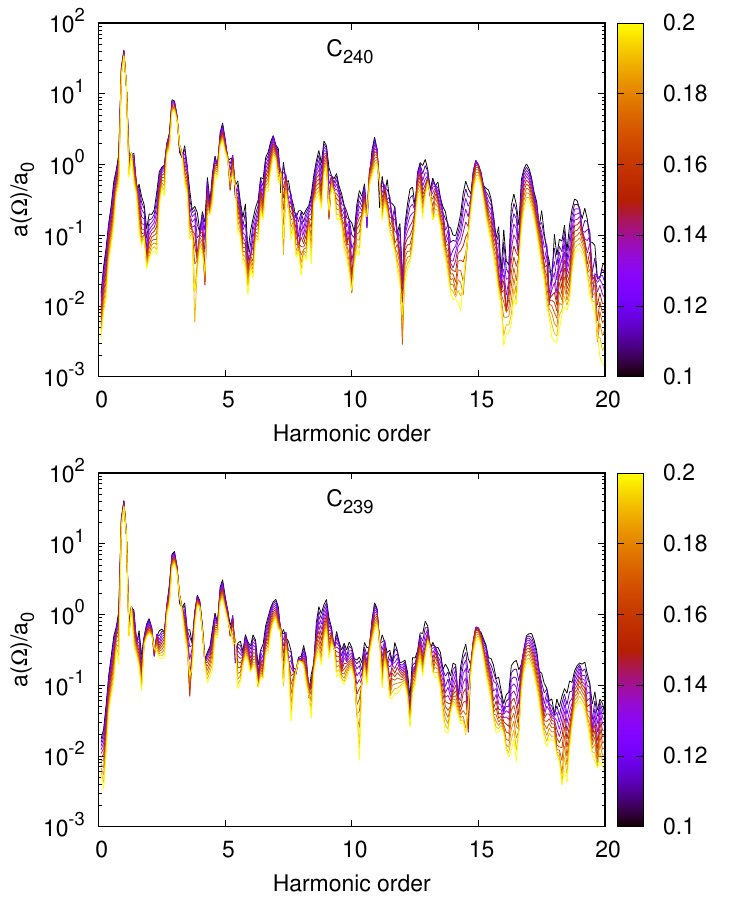}
\caption{The dependencies of the HHG spectra on the relaxation rate
illustrated for C$_{240}$ (top) and C$_{239}$ (bottom). The spectra are
shown for EEI energy: U = 6 eV. The pump wave frequency is $\protect\omega %
=1.5\ \mathrm{eV}/\hbar $. The wave amplitude is taken to be $E_{0}=0.5\ 
\mathrm{V/\mathring{A}}$ The color bar shows the relaxation rate in $\ 
\mathrm{eV}/\hbar $.}
\end{figure}

As have been seen from Fig. 1, the position of molecular excitonic lines and
relative intensities depend on EEI. It is also expected HHG yield change due
to EEI. The latter is shown in Fig. 8, where the HHG spectra in the
strong-field regime for different EEI energies are shown for fullerene C$%
_{240}$ molecule. The similar picture we have for C$_{239}$ molecule. As is
seen, HHG yield strongly depends on the EEI energy. The inclusion of the
Coulomb interaction leads to two noteworthy characteristics in the HHG
spectra: (a) the most prominent feature is a substantial increase in the HHG
signal by several orders of magnitude near the cutoff regime compared to the
case of free quasiparticles. (b) The cutoff frequency is significantly
enhanced. The significant enhancement in the HHG signal can be explained by
the strong modification of hopping integrals (\ref{tau}) and the resulting
level dressing due to the mean field effect. This observation gains further
support from the noticeable prominence of these features in the case of the
giant fullerene C$_{240}$, in stark contrast to the behavior observed in C$%
_{60}$ molecule \cite{avetissian2021high}. Another notable aspect of the HHG
signals in giant fullerene molecules is their dependence on the size of the
molecule. The HHG signals per particle for C$_{240}$ and C$_{60}$ are
compared in Fig. 9. As demonstrated, there is a significant increase in the
HHG signal for C$_{240}$ molecule, a result also observed for C$_{70}$
molecule according to previous studies \cite{avetissian2021high}. This
enhancement may be attributed to the density of states, which is indirectly
reflected in Fig. 1 via the absorption spectra. The inset in Fig. 9 shows
the linear absorption spectrum for C$_{60}$ molecule obtained in the same
way, as in Fig. 1. This figure reveals that C$_{240}$ molecule has
substantially more transition channels than C$_{60}$ one.

\begin{figure}[tbp]
\includegraphics[width=0.42\textwidth]{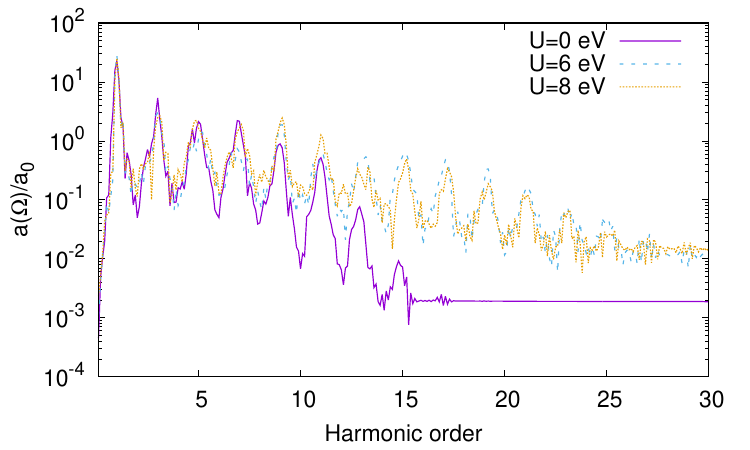}
\caption{The comparision of HHG signals for C$_{240}$ at different EEI
energies. The pump wave frequency is $\protect\omega =1.2\ \mathrm{eV}/\hbar 
$ and wave amplitude is 0.5 V/A. The relaxation rate is set to $\hbar 
\protect\gamma =0.1\ \mathrm{eV}$.}
\end{figure}

\begin{figure}[tbp]
\includegraphics[width=0.42\textwidth]{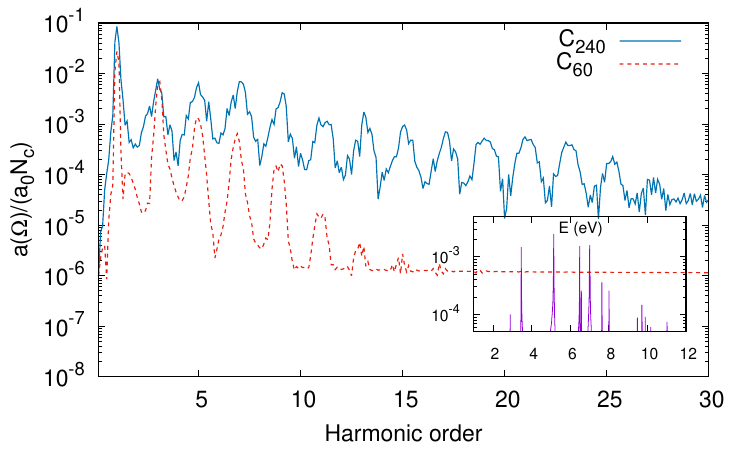}
\caption{The comparision of HHG signals per particle for C$_{240}$ and C$%
_{60}$. The pump wave frequency is $\protect\omega =1.1\ \mathrm{eV}/\hbar$
and wave amplitude is 0.5 V/A. The relaxation rate is set to $\hbar \protect%
\gamma =0.1\ \mathrm{eV}$. The inset shows the linear absorption spectrum
for C$_{60}$ obtained in the same way as in Fig. 1.}
\end{figure}

Finally, note that within the scope of described methodology we have
explored the correlation between the cutoff frequency and the intensity of a
pump wave by analysing the HHG spectra for various intensities. The
relationship between the HHG spectra and the amplitude of the wave field for
both giant molecules is visually represented in Fig. 10. This figure
prominently illustrates the nonlinear connection between the pre-plateau
harmonics and the amplitude of the pump wave. The analysis of obtained
results reveals that for high intensities, the positions of the cutoff
harmonics can be adequately described by scaling with the square root of the
field strength amplitude. The solid lines superimposed on the density plot
in Fig. 10, represent envelopes ($\sim \sqrt{E_{0}}$) that determine the
positions of the cutoff harmonics. Notably, it is evident that these
envelopes provide a reasonably accurate approximation for the cutoff
harmonics for a large field strengths.

\begin{figure}[tbp]
\includegraphics[width=0.42\textwidth]{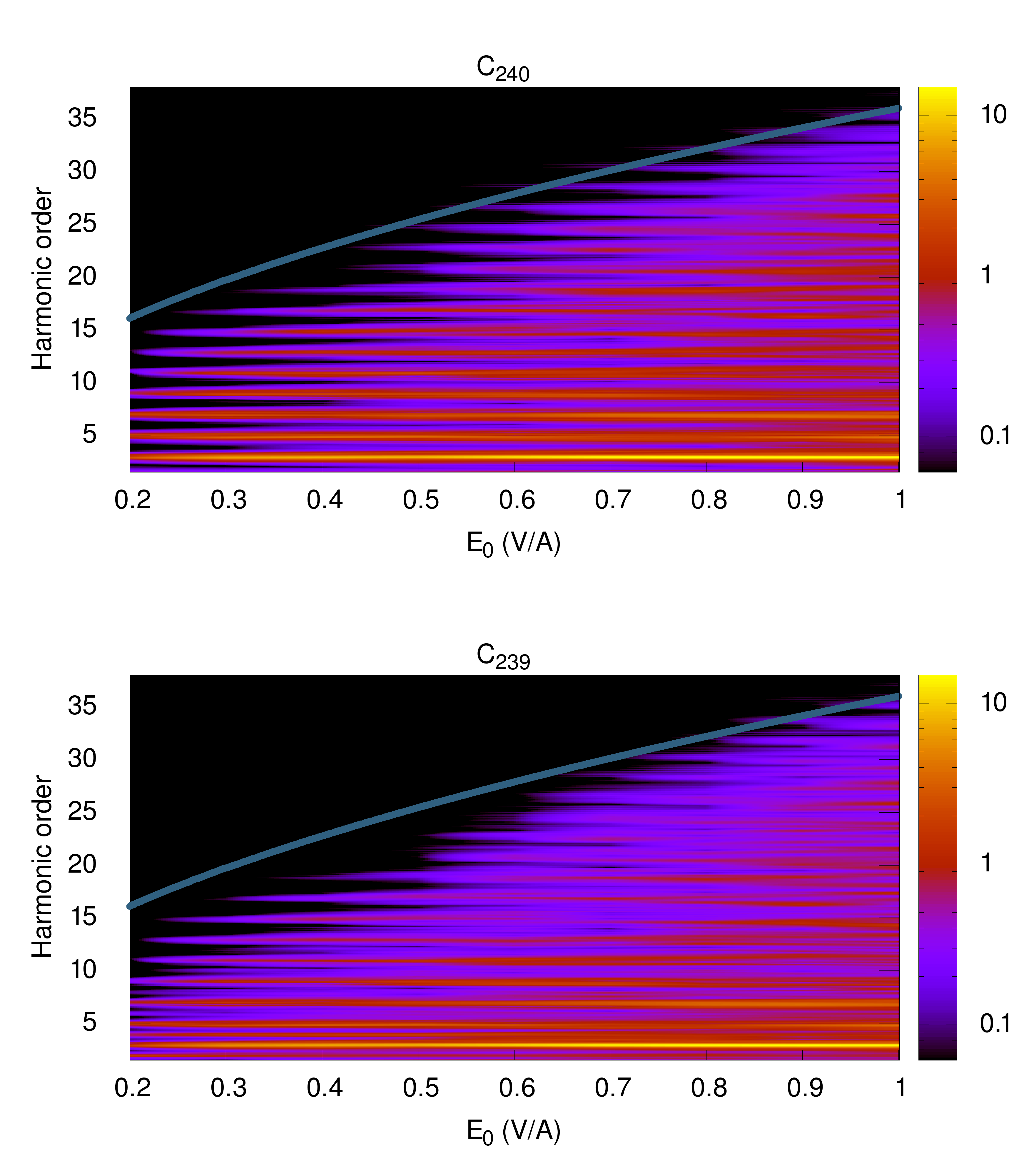}
\caption{The dependencies of the HHG spectra on the wave field amplitude is
illustrated for C$_{240}$ (top) and C$_{239}$ (bottom) using the normalized
dipole acceleration Fourier transformation, $a\left( \Omega \right) /a_{0}$
(color bar), plotted on a logarithmic scale. The spectra are shown for EEI
energy: U = 6 eV. The pump wave frequency is $\protect\omega =1.5\ \mathrm{eV%
}/\hbar $.}
\end{figure}

\section{Conclusion}

We have done an extensive exploration of the highly nonlinear optical
response of giant fullerene molecules, with a particular emphasis on C$_{240}
$, which possesses the characteristic icosahedral point group symmetry often
encountered in such molecular systems. To disclose the complete physical
picture of HHG process on giant fullerene molecules with the mentioned
icosahedral symmetry, we have also investigated a vacancy-defected molecule,
C$_{239}$. Our investigation employed consistent quantum/analytic and
numerical calculation of the HHG spectra using a mean-field methodology that
rigorously accounts for long-range many-body Coulomb interactions too.
Through the solution of the evolutionary equations governing the
single-particle density matrix we have disclosed resonant effects within the
HHG spectra and have demonstrated the fundamental role of Coulomb
interaction in shaping the intensities of the harmonics. A significant
enhancement in HHG yield, as compared with fullerene molecule C$_{60}$, has
been established. Moreover, our research has elucidated that the presence of
a single vacancy, causing the breakdown of icosahedral symmetry, stimulates
the appearance of pronounced even-order harmonics. In terms of the
dependence of the cutoff harmonics on the intensity of the wave field, we
have established that this relationship can be approximated with greate
accuracy by scaling with the square root of the amplitude of a pump wave
strength.

\begin{acknowledgments}
The work was supported by the Science Committee of Republic of
Armenia, project No. 21AG-1C014.
\end{acknowledgments}

\bibliographystyle{apsrev4-2}
\bibliography{bibliography}

\end{document}